\documentclass[traditabstract]{aa}
\usepackage{natbib}
\bibpunct{(}{)}{;}{a}{}{ , }
\usepackage[dvipdfm]{graphicx}
\usepackage[dvips]{color} 
\usepackage{amsmath}
\usepackage{setspace}
\usepackage{booktabs}
\usepackage{txfonts}

\begin{document}

\title{An improved cosmic crystallography method to detect holonomies in flat spaces}
\author{Hirokazu Fujii  \& Yuzuru Yoshii}
\institute{Institute of Astronomy, School of Science, University of Tokyo, 2-21-1, Osawa, Mitaka, Tokyo 181-0015, Japan}
\date{Received 19 January 2011 / Accepted 2 March 2011}

\abstract{
A new, improved version of a cosmic crystallography method for constraining cosmic topology is introduced. Like the circles-in-the-sky method using CMB data, we work in a thin, shell-like region containing plenty of objects. Two pairs of objects (quadruplet) linked by a holonomy show a specific distribution pattern, and three filters of \emph{separation, vectorial condition}, and \emph{lifetime of objects} extract these quadruplets. Each object $P_i$ is assigned an integer $s_i$, which is the number of candidate quadruplets including $P_i$ as their members. Then an additional device of $s_i$-histogram is used to extract topological ghosts, which tend to have high values of $s_i$. 
In this paper we consider flat spaces with Euclidean geometry, and the filters are designed to constrain their holonomies. As the second filter, we prepared five types that are specialized for constraining specific holonomies: one for translation, one for half-turn corkscrew motion and glide reflection, and three for $n$-th turn corkscrew motion for $n=4, 3,$ and 6. {Every multiconnected space has holonomies that are detected by at least one of these five filters. }
Our method is applied to the catalogs of toy quasars in flat $\Lambda$-CDM universes whose typical sizes correspond to $z\sim 5$. With these simulations our method is found to work quite well. { These are the situations in which type-II pair crystallography methods are insensitive because of the tiny number of ghosts. Moreover, in the flat cases, our method should be more sensitive than the type-I pair (or, in general, $n$-tuplet) methods because of its multifilter construction and its independence from $n$.}}

\keywords{cosmology: theory - cosmology: large scale structure of Universe}

\titlerunning{An improved crystallography method for flat spaces}
\authorrunning{H. Fujii \& Y. Yoshii}

\maketitle

\section{Introduction}
The shape of our space is one of the greatest and the oldest questions in human history. Ancient people tried to answer the question mythologically or philosophically, but in the present era, we are ready to answer it scientifically. When we consider the universe as a 4-manifold consisting of 3-space and 1-time, answering the question is translated into determining the local geometry and the global topology of our  {Universe}.

Local geometry is described by Einstein's General Relativity (GR). The metric of space-time plays a fundamental role, and the assumption of the cosmological principle, which states that our  {Universe} is (locally) homogeneous and isotropic, leads us to a famous Friedman-Lema\^ itre-Robertson-Walker (FLRW) metric,
\begin{equation}
ds^2=-c^2 dt^2 +R(t)^2 \Bigl [ \frac{dr^2}{1-kr^2} + r^2 (d \theta ^2 + \sin ^2 \theta d \phi ^2 ) \Bigr ],
\end{equation}
where $k$ is the normalized curvature of space, such as $k= +1$ (spherical geometry), 0 (Euclidean geometry), or $-1$ (hyperbolic geometry). Together with Einstein's equations, curvature is related to the average energy density of the universe and then to other physical quantities. Recent observations favor a $\Lambda$-CDM universe with curvature $k\simeq0$ (e.g. $\Omega_{\mathrm{tot}}=1.0050^{+0.0060}_{-0.0061}$ from \begin{it}WMAP\end{it}+BAO+SN data, by Hinshaw et al. 2009), suggesting that our  {Universe} is one of the flat spaces with Euclidean geometry. 

Global topology, on the other hand, has no reliable physical theories to describe it, so  { the easiest approach is to} constrain it mathematically through direct observations. This situation is analogous to the case of Carl  {Friedrich} Gauss who had to measure the angles of the large triangle formed by three peaks of mountains to know which geometry describes our space, since he did not know GR and modern cosmology.

Moreover, it is true that the global topology of the universe has been a relatively less popular concept than the local geometry. A pioneering work on cosmic topology was done by Ellis (1971). Other early works include Sokolov and Shvartsman (1974), Fang and Sato (1985), Gott (1980), Fangundes (1983), and so on. This theoretical or observational research has not  {attracted much notice} due to the lack of observational data. In the past two decades, however, we have seen tremendous progress in this field, along with progress in observational techniques. Specifically, a possibility of the Poincar\'e dodecahedral space topology suggested by Luminet et al. (2003) was a breakthrough in cosmic topology.  This field recieves more and more interest these days, not only from theorists but also from observational astronomers. The overall topology of the  {Universe} is now becoming one of the major concerns in astronomy and cosmology.

In modern cosmology with an FLRW metric, already described above, we have three geometries  { (scalar curvatures) according to their sign}. For each geometry, there is only one type of space with a simply connected topology, namely, 3-sphere $\mathbb{S}^3$, 3-Euclidean space $\mathbb{E}^3$, and 3-hyperbolic space $\mathbb{H}^3$, for $k=+1, 0,$ and $-1$, respectively. It is possible to construct a space that is locally indistinguishable from the simply connected one, i.e., having the same  { curvature but with a different topology. } It is not simply connected, but multiconnected. 

We give a brief review of multiconnected spaces below. Detailed treatments are found in various reviews (e.g. Lachi\`eze-Rey and Luminet 1995). A multiconnected space $M$ is a quotient space of the simply connected space $U$ with the same geometry, by a holonomy group $\Gamma$, where a holonomy is an isometry on $U$ without any fixed points (except for the identity). Hence, $M=U/\Gamma$ can be  {imagined} as $U$ tiled by polyhedra identified by holonomies $\gamma \in \Gamma$, possessing  {repeated} structures. This polyhedron, called a fundamental cell, is a $2K$-polyhedron whose $K$ pairs of faces are glued mathematically by holonomies. The definition of fundamental cell is not unique, and in this paper by fundamental cell we mean the Dirichlet domain seen from the ``center" of the universe, $D(\vec x_0)=\{\vec x \in U \ | \ \forall \gamma \in \Gamma, |\vec x - \vec x_0| \leq |\vec x -\gamma \vec x _0| \}$. In some space that is globally inhomogeneous, we can define a center, and the fundamental cells for 17 multiconnected flat spaces are found in Figure~1. No two points in a Dirichlet domain  {can} be linked by any holonomies; in this sense a Dirichlet domain represents the whole universe. In a globally homogeneous space, the shape of the Dirichlet domain is independent of the observer's position, since it is always the same as that of the fundamental cell. In a globally inhomogeneous space, on the other hand, in general we do not stand at the center $\vec x_0 $, so the observed Dirichlet domain $D(\vec x_{\mathrm{obs}})$ can vary from the fundamental cell $D(\vec x_{\mathrm{0}})$.

Because of the  {repeated} structures given by holonomies, we can observe multiple copies (ghost images) of single objects in a multiconnected space that is small compared to the observed region, and most methods for constraining cosmic topology are based on this prediction. If we find such copies in the sky, it suggests that our  {Universe} is multiconnected, while finding no ghosts indicates that our  {Universe} is simply connected, at least within our sight, so the lower limit to the size of our  {Universe} is obtained.

There are mainly two methods for constraining cosmic topology. One is to use the  {3D} distributions of astronomical objects such as galaxies, quasars, and galaxy clusters,  {often} called \begin{it}cosmic crystallography\end{it} method, treated in detail in this paper.  {Use of the terminology of ``cosmic crystallography" varies among authors, and in this paper we use it as the generic name for the 3D data methods. } {This method can be classified into two types according to the type of topological signatures of spatial patterns they search for. Those which search for type-I pairs (or, in general, $n$-tuplets), $(\vec x,\vec y)$ and $(\gamma  \vec x, \gamma \vec y)$, with the relation of $|\vec x- \vec y | = |\gamma \vec x - \gamma \vec y | $, have been proposed by, } {e.g. Roukema (1996) and Uzan et al. (1999). Roukema (1996) has developed a method of directly searching for two $n$-tuplets whose spatial distributions are the same within a tolerance, while Uzan et al. (1999) have developed a method to count up the number of same separations seen in a catalog, which is expected to be larger in a small universe. } {Type-I pairs are found in every multiconnected space since any holonomies are isometries that preserve distance. The others, which search for type-II pairs, $(\vec x, \gamma \vec x)$ and $(\vec y, \gamma \vec y)$ with the relation of $|\vec x - \gamma \vec x|=|\vec y- \gamma \vec y|$, have been proposed by Lehoucq et al. (1996) and Marecki et al. (2005). } {Type-II pairs are only found in spaces where some of their holonomies are Clifford translations, which translate all the points by the same distance. } {Lehoucq et al. (1996) have developed a method to detect Clifford translations using pair-separation histograms. The existence of such holonomies can be found as sharp spikes in histograms at the separations corresponding to their translating distances. Marecki et al. (2005) have improved the method by also using vectorial condition, not just separation. }  This ``type-I and type-II pair" terminology was first introduced in Lehoucq et al. (1999).

The other is to use  {the 2D cosmic microwave background (CMB) maps. Starobinsky (1993), and Stevens et al. (1993) have simulated CMB maps and angular power spectra of temperature fluctuations, assuming 3-torus topology, and compared them with the observation by COBE satellite to obtain lower limits of the size of our Universe. Such simulations have been carried out by various authors (e.g. Luminet et al. 2003; Uzan et al. 2004; Riazuelo et al. 2004; Aurich et al. 2008; Aurich \& Lustig 2010). These simulations and the observed quadrupole suppression suggest that we live in a small universe. There is another, direct observational method to detect topological signatures in CMB maps }called \begin{it}circles-in-the-sky \end{it}method (Cornish et al. 1998). This method uses pairs of circles with the same temperature fluctuation pattern, which may be the  intersections of the last-scattering surface (LSS) and the observer's Dirichlet domain. Cosmic crystallography method lost its popularity after the \begin{it}WMAP\end{it} data release, since these CMB-based methods can constrain more topologies than the former, simply because the CMB data covers the larger region. Several authors have searched for these circles and obtained diverse results. Some claim  that most of nontrivial topologies are ruled out (Cornish et al. 2004; Key et al. 2007). They have searched for antipodal or nearly antipodal pairs of circles in the \begin{it}WMAP\end{it} map, and found no such circles to obtain the lower limit of the cell size as $\sim$ 24 Gpc. This constraint cannot be applied to those spaces whose matched circles can be highly deviated from antipodal. Other authors claim that they have found the hints of multiconnected spaces using improved versions of the circles-in-the-sky method (e.g. Roukema et al. 2008; Aurich 2008), which is inconsistent with the former claim. This disagreement suggests there are  methodological problems, which motivates us to revisit the cosmic crystallography method.

 {As mentioned above, } cosmic crystallography method has many versions since the original ones (Lehoucq et al. 1996; Roukema 1996) were introduced. These existing versions, however, are no longer useful in universes that are comparable to the observed region in size due to the lack of the topological ghosts, which makes them unrealistic for practical application.  { This indicates the need to construct a more sensitive method, and in this paper we present such a new method that assumes Euclidean geometry.} We acknowledge that we have abandoned some generalities so as to enhance signal, but it is not serious because we have full mathematical knowledge about flat topologies. This paper focuses on methodology, and the simulated catalogs of ``toy quasars" used here ignore some actual physics, such as spatial correlations and cycles of activity, since these effects do not affect the general results. 

In section 2 we review the 18 3D spaces described by Euclidean geometry and their holonomies. In section 3 we describe the basics of the cosmic crystallography method and our new techniques for contrasting topological signals. To see the success of our improvements, we apply our method to catalogs of toy quasars. In section 4 we describe our catalogs and models of topologies to be tested. In section 5 we show the results and give some discussions. Conclusions are given in section 6.  {All calculations throughout this paper are done in comoving space.}

\

\section{Flat spaces and their holonomies}

\subsection{Classification of the 3D flat manifolds}
Throughout this paper we are interested in spaces with zero curvature, such as those, described by Euclidean geometry, as suggested by many observations (e.g. $\Omega_{\mathrm{tot}}=1.0050^{+0.0060}_{-0.0061}$ from \begin{it}WMAP\end{it}+BAO+SN data, by Hinshaw et al. 2009). We therefore review a complete set of the 18 flat manifolds with three dimensions. There are no other such topologies mathematically (Nowacki 1934), meaning that our  {Universe} is one of them, if $\Omega_{\mathrm{tot}}$ is exactly equal to unity. The classification is summarized in Table \ref{table1}.

\begin{table*}[!htb]
\centering
\caption{18 3D flat spaces.}
\begin{footnotesize}
\begin{tabular}{c|c|ccc} \hline \hline
 Symbol & Name & Compact & Orientable & Homogeneous \\ \toprule
 $E_1$ & 3-torus & yes & yes & yes \\
 $E_2$ & half-turn space & yes & yes & no\\
 $E_3$ & quarter-turn space & yes & yes & no\\
 $E_4$ & third-turn space & yes & yes & no \\
 $E_5$ & sixth-turn space & yes & yes & no\\
 $E_6$ & Hantzsche-Wendt space & yes & yes & no\\
 $E_7$ & Klein space & yes & no & no\\
 $E_8$ & Klein space with horizontal flip & yes & no & no\\
 $E_9$ & Klein space with vertical flip & yes & no & no\\
 $E_{10}$ & Klein space with half turn & yes & no & no\\
 $E_{11}$ & chimney space & no & yes & yes \\
 $E_{12}$ & chimney space with half turn & no & yes & no \\
 $E_{13}$ & chimney space with vertical flip & no & no & no \\
 $E_{14}$ & chimney space with horizontal flip & no & no & no \\
 $E_{15}$ & chimney space with half turn and flip & no & no & no \\
 $E_{16}$ & slab space & no & yes & yes \\
 $E_{17}$ & slab space with flip & no & no & no \\
 $E_{18}$ &  {simply connected space $\mathbb{E}^3$} & no & yes & yes \\ \bottomrule
 \end{tabular}
 \end{footnotesize}
 \label{table1}
\end{table*}

According to this table, most of the spaces are globally inhomogeneous. The shapes of their  Dirichlet domains depend on the observer's location, and especially their linked faces can be highly deviated from antipodal. Moreover, $E_6, E_7,E_8,E_9,$ and $E_{10}$ have nonantipodal faces linked by holonomies that are independent of the observer's location in the universe (e.g. Riazuelo et al. 2004; Mota et al. 2010). Negative results to a multiconnected  {Universe} obtained by Cornish et al. (2004) and Key et al. (2007) using \begin{it}WMAP\end{it} data are based on the searches for antipodal or nearly antipodal  pairs of circles, so their constraints $L \gtrsim 24$ Gpc cannot be applied to these spaces anyway. It means that these spaces remain as possible models of our  {Universe}, and they may be detectable by a cosmic crystallography method. 

\begin{figure}[!htb]
\centering
\resizebox{\hsize}{!}{\includegraphics{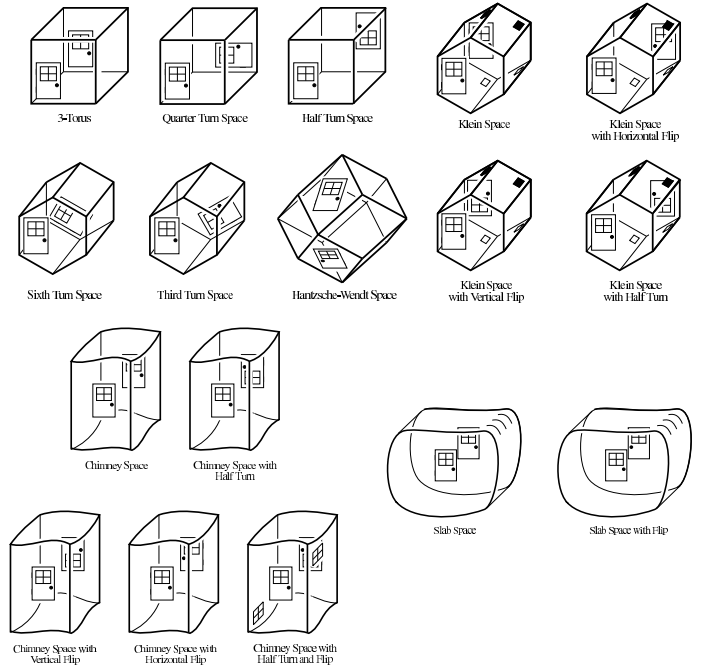}}
\caption{Illustration of fundamental cells for the 17 multiconnected spaces with Euclidean geometry. The chimney space family has two compact dimensions, while the slab space family only has one compact dimension. The others have three compact dimensions; they are compact. The simply connected space $  \mathbb{E}^3$ is not shown here. {\textrm{Courtesy of Jeffrey Weeks and Adam Weeks Marano. This figure was first published in Cipra (2002).}}}
\label{figure1}
\end{figure}

The 17 manifolds $E_1, \cdots, E_{17}$ are all multiconnected, whose fundamental cells are illustrated in Figure \ref{figure1}. The doors or other marks on the faces represent the means for topological gluing. The unmarked faces are glued straight across by parallel translation, except for the Hantzsche-Wendt space, where six pairs of faces are glued in the same way as the doored one.

\

\subsection{Elemental transformations in flat spaces}
 {Hereafter, we use a 4D coordinate system $(w,x,y,z)$ to consider the 3-plane $w = 1$  as the simply connected Euclidean space, so a point $(x,y,z)$ in $\mathbb{E}^3$ is represented as $(1, x, y, z)$. } With this notation, transformations in 3-Euclidean geometry can be written as $4 \times 4$ matrices. We introduce nine elemental transformations below. They are not commutative, but generators of the holonomy group of each space can be derived by their products. In this sense, they are elemental.

\

\noindent \begin{it}1. \ Parallel translation\end{it}
\begin{footnotesize}
\[ T _x(L) = \left( \begin{tabular}{cccc}1 & 0 & 0 & 0 \\ $L$ & 1 & 0 & 0 \\ 0 & 0 & 1 & 0 \\ 0 & 0 & 0 & 1  \end{tabular} \right ), \ T_y(L) = \left (\begin{tabular}{cccc} 1 & 0 & 0 & 0 \\ 0 & 1 & 0 & 0 \\ $L$ & 0 & 1 & 0 \\ 0 & 0 & 0 & 1  \end{tabular} \right ),\]
\[ T_z(L) = \left( \begin{tabular}{cccc} 1 & 0 & 0 & 0 \\ 0 & 1 & 0 & 0 \\ 0 & 0 & 1 & 0 \\ $L$ & 0 & 0 & 1  \end{tabular} \right ).\]\end{footnotesize}

These are the generators of parallel translation along three coordinate axes, by a distance $L$. Parallel translation induces no global inhomogeneity because it operates all points equally.

\

\noindent \begin{it}2. \ Rotation\end{it}
\begin{scriptsize}
\[ O _x(\theta) = \left( \begin{tabular}{cccc}1 & 0 & 0 & 0 \\ 0 & 1 & 0 & 0 \\ 0 & 0 & $\cos \theta$ & $-\sin \theta$ \\ 0 & 0 & $\sin \theta$ & $\cos \theta$  \end{tabular} \right ), \ O_y(\theta) = \left (\begin{tabular}{cccc} 1 & 0 & 0 & 0 \\ 0 & $\cos \theta$ & 0 & $\sin \theta$ \\ 0 & 0 & 1 & 0 \\ 0 & $-\sin \theta$ & 0 & $\cos \theta$  \end{tabular} \right ),\]\end{scriptsize}
\begin{scriptsize}
\[O_z(\theta) = \left( \begin{tabular}{cccc} 1 & 0 & 0 & 0 \\ 0 & $\cos \theta$ & $-\sin \theta$ & 0 \\ 0 & $\sin \theta$ & $\cos \theta$ & 0 \\ 0 & 0 & 0 & 1  \end{tabular} \right ).\]\end{scriptsize}

These are the generators of rotation about three coordinate axes, by an angle $\theta$.  {Rotation is a globally inhomogeneous isometry. } For example, a point near the $x$-axis is translated to a nearby point, while a point distant from the $x$-axis is translated to a distant point, by the same rotation $O_x(\theta)$. 

\

\noindent \begin{it}3. \ Reflection\end{it}
\begin{footnotesize}
\[ R _x = \left( \begin{tabular}{cccc}1 & 0 & 0 & 0 \\ 0 & -1 & 0 & 0 \\ 0 & 0 & 1 & 0 \\ 0 & 0 & 0 & 1  \end{tabular} \right ), \ R_y = \left (\begin{tabular}{cccc} 1 & 0 & 0 & 0 \\ 0 & 1 & 0 & 0 \\ 0 & 0 & -1 & 0 \\ 0 & 0 & 0 & 1  \end{tabular} \right ),\] 
\[ R_z= \left( \begin{tabular}{cccc} 1 & 0 & 0 & 0 \\ 0 & 1 & 0 & 0 \\ 0 & 0 & 1 & 0 \\ 0 & 0 & 0 & -1  \end{tabular} \right ).\]\end{footnotesize}

These are the generators of reflection about three planes perpendicular to coordinate axes, e.g. $R_x$ is a reflection about the $y$-$z$ plane.  {Reflection is a globally inhomogeneous isometry.} For example, a point near the $y$-$z$ plane is translated to a nearby point, while a point distant from the $y$-$z$ plane is translated to a distant point, by the same reflection $R_x$. Moreover,  {the nonorientability of a nonorientable space comes from reflection. }

\subsection{Neighboring holonomies}

With these elemental transformations, generators of the holonomy groups of each space are given as their multiplications. Specifically, neighboring holonomies, i.e., holonomies that link an object in the observer's Dirichlet domain to its topological copies in neighboring cells, are also written as products of elemental transformations as shown in Table \ref{table2}. In sections 4 and 5 we consider the most conservative cases in which neighboring holonomies are barely detectable.

Along with Table 1, it can be seen that the spaces whose holonomy groups contain reflection or rotation are globally inhomogeneous, and those whose holonomy groups contain reflection are not orientable.

\begin{table*}[!htb]
\centering
 \caption{Neighboring holonomies of each topology represented as products of elemental  transformations.}
\begin{footnotesize}
\begin{tabular}{c|l} \hline \hline
 Space symbol & Neighboring holonomies  \\ \toprule
 $E_1$ & $T_x(\pm L_1), \ T_y(\pm L_2), \ T_z(\pm L_3)$ \\
 $E_2$ & $T_x(\pm L_1), \ T_y(\pm L_2), \ T_z(\pm L_3)O_z(\pi)$\\
 $E_3$ & $T_x(\pm L_1), \ T_y(\pm L_1), \ T_z(\pm L_2)O_z(\pm \pi/2)$\\
 $E_4$ & $T_x(\pm L_1)T_y(\pm \sqrt{3}L_1), T_x(\pm 2L_1), \ T_z(\pm L_2)O_z(\pm 2\pi/3)$ \\
 $E_5$ & $T_x(\pm L_1)T_y(\pm \sqrt{3}L_1), T_x(\pm 2L_1), \ T_z(\pm L_2)O_z(\pm \pi/3)$ \\
 $E_6$ & $T_x(\pm L_1)T_y(\pm L_2)O_x(\pi), \ T_y(\pm L_2)T_z(\pm L_3)O_y(\pi), \ T_z(\pm L_3)T_x(\pm L_1)O_z(\pi)$ \\
 $E_7$ & $T_x(\pm L_1)T_y(\pm L_2)R_y, \ T_x(\pm 2L_1), \ T_z(\pm L_3)$ \\ 
 $E_8$ & $T_x(\pm L_1)T_y(\pm L_2)R_y, \ T_x(\pm 2L_1), \ T_z(\pm L_3)R_x$ \\
 $E_9$ & $T_x(\pm L_1)T_y(\pm L_2)R_y, \ T_x(\pm 2L_1), \ T_z(\pm L_3)R_y$ \\
 $E_{10}$ & $T_x(\pm L_1)T_y(\pm L_2)R_y, \ T_x(\pm 2L_1), \ T_z(\pm L_3)O_z(\pi)$ \\
 $E_{11}$ & $T_x(\pm L_1), \ T_y(\pm L_2)$ \\
 $E_{12}$ & $T_x(\pm L_1), \ T_y(\pm L_2)O_y(\pi)$ \\
 $E_{13}$ & $T_x(\pm L_1), \ T_y(\pm L_2)R_z$ \\
 $E_{14}$ & $T_x(\pm L_1), \ T_y(\pm L_2)R_x$ \\
 $E_{15}$ & $T_x(\pm L_1)R_z, \ T_y(\pm L_2)O_y(\pi)$ \\
 $E_{16}$ & $T_z(\pm L_1)$ \\
 $E_{17}$ & $T_z(\pm L_1)R_x$ \\
 $E_{18}$ & none \\ \bottomrule
 \end{tabular}
 \end{footnotesize}
 \label{table2}
\end{table*}

Neighboring holonomies can be classified into six types: half-turn corkscrew motion, quarter-turn corkscrew motion, third-turn corkscrew motion, sixth-turn corkscrew motion, and glide reflection. Holonomies that come from different topologies, but belong to the same type, are indistinguishable by themselves. For example, if we observe a pair of antipodal translations, we live in one of the spaces that have translation as neighboring ones, but we cannot specify in which space we live.

In the following, the spaces in the parentheses are those that have the type of holonomies as neighboring ones.

\

\noindent \begin{it}1. \ Translation\end{it} 

 ($E_1,E_2,E_3,E_4,E_5,E_7,E_8,E_9,E_{10},E_{11},E_{12},E_{13}, E_{14},E_{16}$)

Translation induces antipodal faces of the Dirichlet domain that are glued straight across. Translation is subdivided into three types of one-pair translation (type I), two-pair translation (type II), and three-pair translation (type III). Two-pair translation only exists in quarter-turn space $E_3$; its four {holonomies} $T_x(\pm L)$ and $T_y(\pm L)$ have the common translating distance $L$,  {so either all four holonomies are detectable or all of them are undetectable.} Three-pair translation, on the other hand, exists in third-turn space $E_4$ and sixth-turn space $E_5$; their six  {holonomies} $T_x(\pm \frac{1}{2}L)T_y(\pm \frac{\sqrt{3}}{2}L)$ and $T_x(\pm L)$ have the common translating distance $L$, so   { either all six holonomies are detectable or all of them are undetectable.} One-pair translation exists in the other spaces.

\

\noindent \begin{it}2. \ Half-turn corkscrew motion  \end{it} \ ($E_2, E_6,E_{10},E_{12},E_{15}$)

Half-turn corkscrew motion is the half-turn followed by parallel translation. Half-turn corkscrew motion is subdivided into two types of those that have rotational axes corresponding to translational directions (type I), and those with  different rotational axes and translational directions (type II). Only $E_6$ has the latter type, while the other four spaces have the former.

\

\noindent \begin{it}3. \ Quarter-turn corkscrew motion \end{it} \ ($E_3$)

Quarter-turn corkscrew motion is the quarter-turn followed by parallel translation. Only $E_3$ has quarter-turn corkscrew motion, so detecting such holonomies suggests that we live in a quarter-turn space.

\

\noindent \begin{it}4. \ Third-turn corkscrew motion \end{it} \ ($E_4$)

Third-turn corkscrew motion is the third-turn followed by parallel translation. Only $E_4$ has third-turn corkscrew motion, so detecting such holonomies suggests that we live in a third-turn space.

\

\noindent \begin{it}5. \ Sixth-turn corkscrew motion \end{it} \ ($E_5$)

Sixth-turn corkscrew motion is the sixth-turn followed by parallel translation. Only $E_5$ has sixth-turn corkscrew motion, so detecting such holonomies suggests that we live in a sixth-turn space.

\

\noindent \begin{it}6. \ Glide reflection \end{it} \ ($E_7,E_8, E_9,E_{10},E_{13},E_{14},E_{15},E_{17}$)

Glide reflection is the reflection followed by parallel translation. They are subdivided into two types of those that have reflectional planes parallel to translational directions (type I) and those with nonparallel reflectional planes and translational directions (type II). The six spaces of  $E_8,E_9,E_{13},E_{14},E_{15},$ and $E_{17}$ have the former type, while the four spaces $E_7,E_8,E_9,$ and $E_{10}$ have the latter.

\

\section{Method}

\subsection{Basic ideas}
Cosmic crystallography is a series of statistical techniques to extract topological information from a given astronomical catalog first proposed by Lehoucq et al. (1996)  and Roukema (1996). The former method is to search for type-II pairs, and the latter is to search for type-I $n$-tuplets (generalization of type-I pairs to $n$ objects).   This type-I, type-II terminology is that of Lehoucq et al. (1999), and is unrelated to the definitions of types in Section 2.3.  In this paper we propose a new version of crystallographic method,  {which is mainly for collecting type-I pairs as in Roukema (1996) and Uzan et al. (1999) and has filters that are related to those of Marecki et al. (2005). } Flat spaces are assumed throughout the paper, but the methodology in this section can be applied to spaces with any curvature.

\begin{figure}[!htb]
\centering
\resizebox{\hsize}{!}{\includegraphics{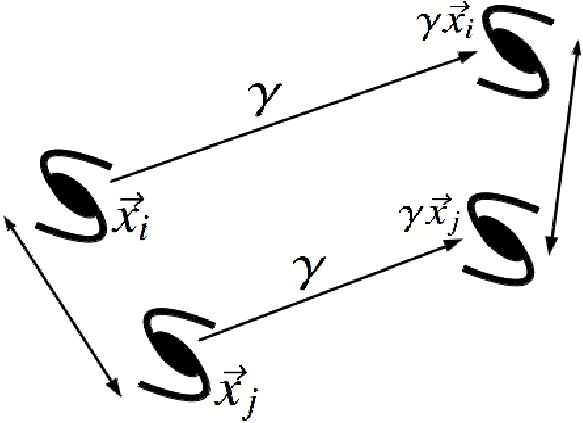}}
\caption{Two pairs of objects linked by a holonomy $\gamma$ ideally have the same separations. This ideal situation breaks down owing to peculiar velocities and observational limits.}
\label{figure2}
\end{figure}

We assume that our  {Universe} has topology $M=\mathbb{E}^3/\Gamma$, where $\mathbb{E}^3$ is a simply connected 3-Euclidean space, and $\Gamma$ is a holonomy group on $\mathbb{E}^3$. Suppose that in our catalog we have $N$ objects $P_1, \cdots, P_N$, whose comoving positions are given by $\vec x_1 , \cdots, \vec x_N$, respectively. We often indicate the object $P_i$ itself by its positonal vector $\vec x_i$ for convenience. Ghosts of two objects $\vec x_i$ and $\vec x_j$ by a holonomy $\gamma \in \Gamma$ are given by $\gamma \vec x_i$ and $\gamma \vec x_j$, respectively. All holonomies are isometries that preserve the distance, so we have
\begin{equation}
|\vec x_i- \vec x_j| = |\gamma \vec x_i- \gamma \vec x_j|,
\end{equation}
independently of the holonomy's other properties (Figure \ref{figure2}).
If we find two pairs $(\vec x_i, \vec x_j)$ and $(\vec x_k, \vec x_l)$ such that $|\vec x_i -\vec x_j|=|\vec x_k - \vec x_l|$, they can be ghosts of each other. We call the two pairs of objects a \begin{it}quadruplet\end{it} hereafter {(notice the difference between the type-I $n$-tuplet and the quadruplet here)}, and the statement above is equal to saying that a quadruplet $[(\vec x_i, \vec x_j),(\vec x_k, \vec x_l)]$ with the relation  $|\vec x_i -\vec x_j|=|\vec x_k - \vec x_l|$ can be topological, and finding such quadruplets indicates the multiconnectedness of our  {Universe}. However, in practice, these separations cannot be exactly the same because of various effects such as peculiar velocities and observational limits, so we search for pairs having separations deviated at most by some positive value $\varepsilon$. Because of this finite  spatial resolution, some nontopological quadruplets stochastically pretend to be topological. The number of such quadruplets $M$ is expected to be proportional to the total number of quadruplets $_N C _4 \times {_4 C} _2 \times \frac{1}{2!}$, so the topological index is given by
\begin{equation}
S=\frac{M}{_N C _4 \times {_4 C} _2 \times \frac{1}{2!}} = \frac{8M}{N(N-1)(N-2)(N-3)}.
\end{equation}
In a small universe, there are real topological quadruplets added to false stochastic ones, so the index $S$ will take a higher value than expected in a simply connected universe. This formulation is slightly different from the CCP method, but their essence is common to both. 

Until now we have only used the general property of holonomies as isometries, i.e. preserving the distance. If a holonomy $\gamma$ is a Clifford translation, which translates all points by the same distance, additional pairs have the same separations:
\begin{equation}
|\vec x_i - \gamma \vec x_i| = |\vec x_j- \gamma \vec x_j|.
\end{equation}
In this case, the topological signal by $\gamma$ is multiplied by a factor of 2. Since we concentrate on spaces with Euclidean geometry, a Clifford translation is a normal parallel translation. 

\subsection{Sophistication}

The existing crystallography method has been verified as useful for constraining cosmic topology, only for topologies whose fundamental cells are so small that the observed region contains no fewer than several tens of cells within it. However, in reality, the fundamental cell of our  {Universe} can be comparable to the observed region in size. Of course it can be larger than the observed region, but in this case no methods can be used, so we do not consider it. For such situations, the crystallography method is no longer valid because the real signal is too weak compared to the false signal, which occurs by chance. It is necessary for us to make the methods more sophisticated in order to decrease the false signal, while keeping the real signal.

\subsubsection{Imitation of the circles-in-the-sky method}
First of all, we propose to concentrate on a thin, shell-like region $r_1<r<r_2$, like the circles-in-the-sky method using LSS. The reason we consider such a region is that we have to reasonably decrease the number of objects, since, roughly speaking, the real signal is proportional to $N^2$, while the false signal is proportional to $N^4$. If we can detect the signal of a space that is comparable to the shell region in size, then any smaller spaces are detectable since they produce stronger signals. In principle, therefore, the larger the region we consider, the more topologies we can constrain. However, the number of objects becomes too small when the region is too large, so we have to find a good balance.

When the region is chosen, then we have to put all the quadruplets through ``filters'' to drop any false stochastic ones. Here we introduce three filters: the first one has already been discussed both in the previous section and in Uzan et al. (1999), but the others are new ones.

\subsubsection{The first filter : separation}

All holonomies are isometries that preserve distance, and we consider the condition
\begin{equation}
||\vec x_i - \vec x_j| - |\vec x_k - \vec x_l ||<\varepsilon
\end{equation}
as the first filter to extract information about topological lensing.  {This corresponds to selecting type-I pairs as in Roukema (1996) and Uzan et al. (1999). } This filter, because of the finite resolution $\varepsilon$, is insufficient when the real signal is weak. Many false quadruplets pretending to be topological hide the real ones. This filter can be used in spaces with any curvature, unlike the other two, since it uses the general property of holonomies.

\subsubsection{The second filter : vectorial condition}

To extract information about  {repeated} structures due to multi-connectedness, we should pay attention not only to separation but also to vectorial properties of holonomies. Vectorial properties are not  common among holonomies, unlike the distance-preserving property, so this filter cannot constrain all holonomies at once. Here we introduce five kinds of filters that constrain their specialized holonomies. These choices are not unique; in particular, we can construct more specialized filters to constrain more specific holonomies, but it induces a drastic increase in free parameters. {In some sense, } {our five filters can be regarded as the generalizations of \emph{bunch-of-pair} (BoP) selection in Marecki et al. (2005), which selects type-II translational pairs that have both the same separations and the same directions, to nontranslational holonomies.}

\

\noindent \begin{it}1 : Translation\end{it}

This filter is for constraining the parallel translation $\gamma=T_x(L_1)T_y(L_2)T_z(L_3)$. When a given quadruplet $[(\vec x_i, \vec x_j),(\vec x_k, \vec x_l)]$ is a $\gamma$-quadruplet, there are two possibilities such that  $\gamma (\vec x_i , \vec x _j)=(\vec x_k, \vec x_l)$ or $\gamma (\vec x_i , \vec x _j)=(\vec x_l, \vec x_k)$. In the former case, we ideally have a relation:
\begin{equation}
x_i-x_j=x_k-x_l, \ y_i-y_j=y_k-y_l, \ z_i-z_j=z_k-z_l.
\end{equation}
In the latter case, on the other hand, we ideally have a relation:
\begin{equation}
x_i-x_j=-(x_k-x_l), \ y_i-y_j=-(y_k-y_l), \ z_i-z_j=-(z_k-z_l).
\end{equation}
We consider these possibilities together to construct a filter:
\begin{equation}
|(x_i -x_j) \pm(x_k - x_l)| < \varepsilon _x,
\end{equation}
\begin{equation}
|(y_i -y_j) \pm(y_k - y_l)| < \varepsilon _y,
\end{equation}
\begin{equation}
|(z_i -z_j) \pm(z_k - z_l)| < \varepsilon _z,
\end{equation}
independent of the translating distance and direction. Note that $\vec \varepsilon=(\varepsilon _x, \varepsilon_y, \varepsilon _z)$ here is different from $\varepsilon$ in the first filter.  {This filter corresponds to the BoP selection in Marecki et al. (2005), but our filter also detects type-I pairs, not only type-II pairs. However, this means that our filter also detects false type-I and type-II pairs, so there seems to be no significant advantages and disadvantages between them.}

\

\noindent \begin{it}2 : Half-turn corkscrew motion and glide reflection\end{it}

This filter is for constraining half-turn corkscrew motion, e.g. $\gamma _1=T_x(L_1)T_y(L_2)$ $T_z(L_3)O_z(\pi)$, and glide reflection, e.g. $\gamma_2=T_x(L_1)T_y(L_2)T_z(L_3)R_z$. For a given quadruplet  $[(\vec x_i, \vec x_j),(\vec x_k, \vec x_l)]$ to be a $\gamma _{n}$-quadruplet, there are two possibilities such that $\gamma_n (\vec x_i , \vec x _j)=(\vec x_k, \vec x_l)$ or $\gamma_n (\vec x_i , \vec x _j)=(\vec x_l, \vec x_k)$.

\noindent In the former case, we ideally have a relation for $n=1$
\begin{equation}
x_i-x_j=-(x_k-x_l), \ y_i-y_j=-(y_k-y_l), \ z_i-z_j=z_k-z_l, 
\end{equation}
and for $n=2$
\begin{equation}
x_i-x_j=x_k-x_l, \ y_i-y_j=y_k-y_l, \ z_i-z_j=-(z_k-z_l).
\end{equation}
In the latter case, on the other hand, we ideally have a relation for $n=1$
\begin{equation}
x_i-x_j=x_k-x_l, \ y_i-y_j=y_k-y_l, \ z_i-z_j=-(z_k-z_l), 
\end{equation}
and for $n=2$
\begin{equation}
x_i-x_j=-(x_k-x_l), \ y_i-y_j=-(y_k-y_l), \ z_i-z_j=z_k-z_l.
\end{equation}
We consider these possibilities together to construct a filter:
\begin{equation}
|(x_i -x_j) \pm(x_k - x_l)| < \varepsilon _x,
\end{equation}
\begin{equation}
|(y_i -y_j) \pm(y_k - y_l)| < \varepsilon _y,
\end{equation}
\begin{equation}
|(z_i -z_j) \mp(z_k - z_l)| < \varepsilon _z.
\end{equation}
Here the rotation and reflection are based on the ``global" directions of the universe about which we primarily have no idea, so in practice we have to find them by changing our choices of coordinate axes (see section 5.2.1 for details).

\

\noindent \begin{it}3 : $n$-th turn corkscrew motion ($n=4,3,6$)\end{it}

These filters are for constraining $n$-th turn corkscrew motion for $n=4,3,$ and 6. It is, for example, a holonomy $\gamma=T_x(L_1)T_y(L_2)T_z(L_3)O_z(2\pi/n)$. As in the previous filters, there are two ways to pass through this filter for a given quadruplet. 

We construct the filters by considering the two possibilities together, say,
\begin{eqnarray}
&|(x_i-x_j)\cos(2\pi/n)-(y_i-y_j)\sin(2\pi/n) \pm (x_k-x_l) | <\varepsilon_x, &\\
&|(x_i-x_j)\sin(2\pi/n)+(y_i-y_j)\cos(2\pi/n) \pm (y_k-y_l) | <\varepsilon_y, &\\
&|(z_i -z_j) \pm (z_k-z_l)| < \varepsilon_z.&
\end{eqnarray}
Corkscrew motion includes rotation based on the global directions, so we have to change our choices of coordinate axes to find it, as already mentioned.

An additional notice here is that in our calculations we do not distinguish two quadruplets $[(\vec x_i, \vec x_j),(\vec x_k, \vec x_l)]$ and $[(\vec x_k, \vec x_l),(\vec x_i, \vec x_j)]$, so the above filters will ignore $\gamma^{-1}$-quadruplets. For this we also use the filters
\begin{eqnarray}
&|(x_i-x_j)\cos(2\pi/n)+(y_i-y_j)\sin(2\pi/n) \pm (x_k-x_l) | <\varepsilon_x, &\\
&|-(x_i-x_j)\sin(2\pi/n)+(y_i-y_j)\cos(2\pi/n) \pm (y_k-y_l) | <\varepsilon_y, &\\
&|(z_i -z_j) \pm (z_k-z_l)| < \varepsilon_z,&
\end{eqnarray}
which are unnecessary for the previous two cases since $\gamma^{-1}$-quadruplets can also be constrained by the same filters. 

\subsubsection{The third filter: lifetime of objects}

Astronomical objects have finite lifetimes $t_{\mathrm{life}}$. Suppose two objects $\vec x_i$ and $\vec x_j$ whose cosmic times are $t_i$ and $t_j$, respectively. In order that they are ghost images of each other,
\begin{equation}
\Delta t_{ij} \equiv |t_i -t_j| < t_{\mathrm{life}}
\end{equation}
is a necessary condition, so a quadruplet [$(\vec x_i,\vec x_j)$,$(\vec x_k, \vec x_l)$] should be dropped unless $\Delta t_{ik}, \Delta t_{jl}<t_{\mathrm{life}}$ or $\Delta t_{il}, \Delta t_{jk}<t_{\mathrm{life}}$.  This very closely corresponds to the redshift filter in Marecki et al. (2005), and the difference is only an expression.  Some preceding studies ignore this effect and use all pairs despite their cosmic times. This is a crucial fault unless $L \ll ct_{\mathrm{life}}$, where $L$ is a characteristic size of the universe. 

\subsubsection{Finishing: classification of objects}
Though the above three filters can drop false stochastic quadruplets while keeping real topological ones, it is possible that the false signal is still strong enough to hide the real signal. For this situation, we introduce an additional technique to contrast real signals. 

Each object $\vec x_i \ (i=1,\cdots,N)$ is assigned an integer $s_i$, the number of final candidate quadruplets that have passed through all filters and include $\vec x_i$ as their members. False signals are stochastic, so the members of false quadruplets are randomly distributed. As a result, the possibility that many final candidates share a single, common object is small, so $s_i$  {rarely} takes a high value. Real signals, on the other hand, all come from ghost pairs. If there are $n \ \gamma$-pairs $(\vec x_i, \gamma \vec x_i), \cdots, (\vec x_j, \gamma \vec x_j)$ where $\gamma$ is not a parallel translation, $s_k \ge n-1$ for each $\vec x_k$, whereas $s_k \ge2(n-1)$ if $\gamma$ is a parallel translation. As a result the histogram of $s_i$ will contrast real topological signals from false stochastic ones. { This technique corresponds to performing the $n$-tuplet searching in Roukema et al. (1996), for all $n$ simultaneously. The CCP method by Uzan et al. (1999) has no such techniques, and is therefore unable to distinguish, for example, one real $n$-tuplet from $n(n-1)/2$ false pairs. }

Through these techniques, candidates of topological ghosts are sampled in a catalog, and their distributions in the sky give us a hint of the topology of our  {Universe}. They trace the topological gluing of the faces of our Dirichlet domain.

\subsection{Spatial resolutions}
In this paper we do not consider any technical uncertainties; i.e., we assume that we have the full-sky, full-redshift,  and 100\% complete catalog with infinite accuracy. However, even in such a situation, peculiar velocity with respect to the comoving flame prevents us from having the ideal catalog where Eq.(2) holds precisely. Effects of peculiar velocities consist of two parts: (i) integrated effect and (ii) instantaneous effect.  {They were first discussed by Roukema (1996), while these terminologies are given by Lehoucq et al. (2000)}. The integrated effect comes from the true motion of objects, and the instantaneous effect comes from the fake positional change due to an  additional redshift by the Doppler effect. To see these effects quantitatively, we consider two objects $P_i$ and $P_j = \gamma P_i$, linked by a holonomy $\gamma$. Their positions are given by $\vec x_i$ and $\vec x_j = \gamma \vec x_i$, respectively, when the object is comoving. In practice, however, the observed positions are slightly different from those. 

First we consider the integrated effect. Denoting the cosmic times of $P_i$ and $P_j$ by $t_i$ and $t_j$, respectively, we obtain
\begin{equation}
\Delta_{\mathrm{int}} \vec x_j  = \int _{t_i} ^{t_j} \frac{\vec v}{a(t)} dt,
\end{equation}
as a motion of the object from $t=t_i$ to $t=t_j$, so it affects the observed position of $P_j$ as an additional term to $\gamma \vec x_i$. Here $\Delta_{\mathrm{int}} \vec x_i = 0$ by definition.

Next we consider the instantaneous effect. Assuming the peculiar velocity $v=|\vec v|$ is time-independent, the extra redshift by the Doppler effect is given by
\begin{equation}
\Delta z_{\mathrm{dop}} = \frac{v_{\mathrm{\parallel}}}{c}(1+z_{\mathrm{cos}} ),
\end{equation}
 {to first order in $v_{\parallel}/c$}, where $v_{\mathrm{\parallel}}$ is a radial component of $\vec v$, and $z_{\mathrm{cos}}$ is a purely cosmological redshift. This extra term affects the observed comoving radial distance of objects via 
\begin{equation}
\Delta _{\mathrm{ins}}r =\frac{c}{H_0} \int _{\frac{1}{1+z_{\text{cos}}+\Delta z_{\text{dop}}}} ^{\frac{1}{1+z_{\text{cos}}}} \frac{da}{\sqrt{\Omega_{m0}a + (1-\Omega_{m0}-\Omega_{\Lambda 0}) + \Omega_{\Lambda 0} a^4}}.
\end{equation}
This one, unlike the integrated effect, affects both $P_i$ and $P_j$ on their observed positions. 

Incorporating these two effects simultaneously, the observed positions of $P_i$ and $P_j$ are given by
\begin{equation}
\vec x_i ^{\mathrm{obs}} = \vec x_i + \Delta _{\mathrm{ins}} r_i \frac{\vec x_i}{|\vec x_i|}
\end{equation}
and
\begin{equation}
\vec x_j ^{\mathrm{obs}} = \gamma \vec x_i + \Delta_{\mathrm{int}} \vec x_j + \Delta _{\mathrm{ins}} r_j \frac{\gamma \vec x_i + \Delta_{\mathrm{int}} \vec x_j}{|\gamma \vec x_i + \Delta_{\mathrm{int}} \vec x_j|},
\end{equation}
respectively. Here we have seen the effects of peculiar velocities on observed positions by which we would fail to recognize real topological quadruplets, which helps us to determine spatial resolutions  $\varepsilon$ in the first filter and $\vec \varepsilon$ in the second filter. They should be low enough to avoid missing the topological quadruplets that are distorted by these effects, however, simultaneously, they should be high enough to decrease the false signal. A balance between them determines $\varepsilon$ and $\vec \varepsilon$ (details are in section 5).

\

\section{Simulations}
\subsection{Toy catalogs of quasars}

To test our new method, on the level of methodology, we generated toy quasar distributions in flat spaces with the standard $\Lambda$-CDM cosmology ($\Omega_\mathrm{m}=0.27, \Omega_{\Lambda}=0.73, H_0=71$ km/sec/Mpc), and applied the method to them. We make complete, full-sky and full-redshift catalogs with infinite accuracy  {(more precisely, double-precision floating-point accuracy)}. As the nature of quasars, we consider the cosmological evolution of comoving density, peculiar velocity, and lifetime, since the most crucial factor is a ratio of the number of topological quadruplets to that of the false quadruplets. The lifetime of quasars concerns the former, and the comoving density and peculiar velocity concern the latter.  The other natures of clustering and cycles of activity are ignored because we focus on methodology here, and these effects are not important at this level. The method itself can be applied to catalogs of any extragalactic objects, such as active galactic nuclei (AGNs), high-$z$ galaxies, and  galaxy clusters. 

The redshift evolution of the comoving density of quasars is taken from Osmer (2004), but we have extrapolated it to $z \sim 7$. We simplified the luminosity evolution of quasars such that they emit radiation with constant luminosity during the fixed duration $t_{\mathrm{life}} = 10^8$ yr. We also simplified the peculiar motion of quasars to move with constant speed $v=500$ km/sec and with randomly chosen directions. 

Every $10^7$ yr from $t=0$ to the present $t=1.37 \times 10^{10}$ yr, new quasars are generated and randomly distributed, and the quasars reaching their lifetime of $10^8$ yr are removed. For simplicity, no spatial correlations are considered in this paper. While the comoving density of quasars used in simulations is common, the actual number of quasars listed in the catalogs is different. This can be viewed as the cosmic variance. 

The topological signal is searched for in a shell region 7.8 Gpc$<r<$8.2 Gpc (corresponding to $4.7<z<5.5$), in which about 2000 quasars are contained. This choice of radial width does not necessarily have significant meanings, but this is close to the limiting size ($z \sim 7$) of the actual distribution of quasars. Hereafter, any catalog mentioned is referred to as a part of the full catalog satisfying 7.8 Gpc$<r<$8.2 Gpc.

\subsection{Models of spaces}
For each topology, the holonomies we target are assumed to have translational parts, e.g., $T_z(L_3)$ for $T_z(L_3)O_z(\pi)$, satisfying
\begin{equation}
\sqrt{L_1^2+L_2^2+L_3^2}=L=16 \ \mathrm{Gpc}.
\end{equation}
which is a limit size that can be detected with our catalog 7.8 Gpc$<r<$8.2 Gpc. If they are detectable, then any other space that is smaller than them is also detectable.  {It would be difficult to make a detection in this case using other 3D methods.}

In each calculation, our policy is ``holonomy first". According to the policy, unless otherwise stated, in each case we consider the most conservative cases in which all holonomies except for the targeted ones are beyond the observed region: they were undetectable. In other words, we consider the  severest case for each topology to detect it. For example, when we target the $z$-parallel translations $T_z(\pm L)$ of 3-torus, we neglect the other translation $T_x(\pm L_1)$ and $T_y(\pm L_2)$ by considering $L_1, L_2 \rightarrow \infty$. When we target the parallel translation of third-turn space, on the other hand, we must simultaneously consider the translation in all directions because of their common translating distance. (see Table 2)

\section{Results and discussions}
 
\subsection{Improvements in detecting topological signal}
Here we show the results in detail for one type of holonomy. Those for the other types are given in the appendix. Two catalogs of toy quasars, one in a slab space with $T_z(\pm L)$ and the other in the  {simply connected }Euclidean space with no holonomies, are compared. Each catalog contains 1980 quasars. Among the 1980 ones in the former catalog, there are six pairs of ghosts, meaning that there are 30 (=${_6 C}_2 \times 2$) topological quadruplets. They are hidden in the large number of false stochastic ones. 
To see the effects of three filters introduced  in section 3, we have increased the number of filters  one by one, and counted the number of quadruplets that have passed through the filters. 

As for spatial resolution, we used $\varepsilon=3.7$ Mpc in the first filter and $\vec \varepsilon=(6.0, 6.0, 6.0)$ in the second filter in units of  Mpc. They are determined as follows.
We generated 10000 topological quadruplets $[(\vec x_i,\vec x_j),(T_z(\pm L) \vec x_i,T_z(\pm L) \vec x_j)]$. The value of $\varepsilon$ was chosen such that 99\% of 10000 quadruplets satisfied the condition $||\vec x_i - \vec x_j|-|T_z(\pm L) \vec x_i - T_z(\pm L) \vec x_j||< \varepsilon$. Then, each component $\varepsilon _x, \varepsilon _y,$ and $ \varepsilon _z$ were determined in the same way as $\varepsilon$.  We denote $\varepsilon '= \mathrm{max} \{ \varepsilon _x, \varepsilon _y, \varepsilon _z \}$ and conservatively substitute $(\varepsilon ',\varepsilon ',\varepsilon ')$ for $\vec \varepsilon$. In principle, we have to set $\varepsilon$ and $\vec \varepsilon$ such that 100\% of the quadruplets satisfy the condition, but it needs very high resolution, and meaninglessly increases the false signal. We balanced them to obtain the values.

The total number of quadruplets is given by
\begin{equation}
M= {_{1980} C} _4 \times {_4 C} _2 \times \frac{1}{2!}=1915375615065.
\end{equation}
This drastically large number of quadruplets are filtered, and the false stochastic ones are dropped as shown in Table \ref{table3}. We can find that the first filter (separation) is insufficient for detecting holonomies when there are only a few ghosts in this case. It suggests that the existing crystallography method like the CCP method is no longer useful for this situation, since the real signal is too weak. The second filter plays an important role in extracting the real signal. The third filter only plays a compensative role here, but will be important in somewhat more delicate situations.
\begin{table}[!htb]
\centering
 \caption{The number of quadruplets ($N_{\mathrm{quad}}$) that have passed through the filters of (1) separation, (2) vectorial condition, and (3) lifetime of objects.}
\begin{tabular}{c|c|c} \hline \hline
Filtering condition & $N_{\mathrm{quad}}$ in the \ \ \ \  & $N_{\mathrm{quad}}$ in the simply  \\ 
& slab space \ \ \ \ & connected space \\ \toprule
no filters & 1915375615065 & 1915375615065 \\
filter (1) & 663853290 & 664887124 \\
filter (1)+(2) & 342 & 323 \\
filter (1)+(2)+(3) & 253 & 239 \\ \bottomrule
 \end{tabular}
 \label{table3}
\end{table}

\begin{figure}[!htb]
\centering
\resizebox{\hsize}{!}{\includegraphics{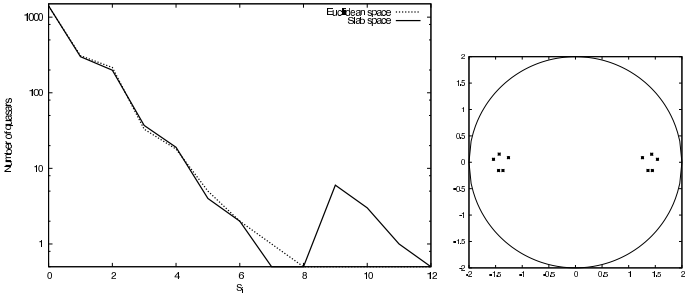}}
\caption{Left: the $s_i$-histograms of the two catalogs with 1980 quasars in the simply connected Euclidean space and in the slab space. $s_i$ is the number of final candidate quadruplets that include the quasar $P_i$. In the slab space, a hill constituted by topological ghosts is clearly seen. Right: The objects with $s_i \geq 7$ in the slab space. It can be seen that they are linked by parallel translation.}
\label{figure3}
\end{figure}

Though the false signal has been decreased successfully, it still hides the real one, so we additionally need the classification of objects (section 3.2.5). The $s_i$-histogram, where $s_i$ is the number of final candidate quadruplets, which include the object $P_i$, for each catalog that is fully filtered is shown in Figure \ref{figure3}. There is a hill that peaks at $s_i=9$, in the histogram for the slab space. It may be constituted by topological ghosts, as expected in section  3.2.5. To check this in Figure \ref{figure3} we display the quasars with $s_i \geq 7$ using the Lambert's azimuthal projection in the $z$-$x$ plane. The pair of five objects are clearly linked by parallel translation, confirming that our space has a translational pair. 

Other examples for this projection are given in Figure \ref{figure4} where we get an advance of the results in Appendix. Quasars with high $s_i$ values, which constitute the hills, are those of a one-pair translation, as in the previous figure, half-turn corkscrew motion (type I), and glide reflection (type I). They clearly trace the way for  topological gluing for each type of holonomies. If the objects that constitute the hills are randomly distributed, it is likely that they are merely stochastic. Moreover, in this stage, even if there are still some false stochastic ones, it is possible to find  {repeated} patterns of real signal with somewhat additional techniques, if exist.

\begin{figure}[!htb]
\centering
\resizebox{\hsize}{!}{\includegraphics{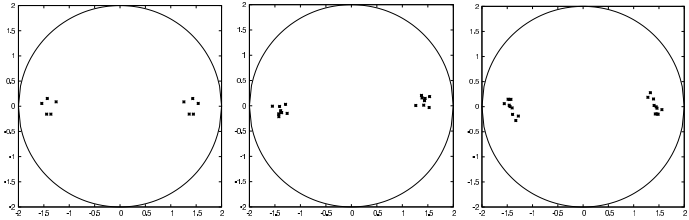}}
\caption{The quasars that constitute the hill in the histogram for each simulation. Left: one-pair translation ($s_i \geq 7$), middle: half-turn corkscrew motion of type I ($s_i \geq 6$), right: glide reflection of type I ($s_i \geq 6$). They clearly trace the topological gluing for each holonomy.}
\label{figure4}
\end{figure}

A relatively large number of quasars at $s_i = 2, 4, $ and 6 in both histograms are due to a condition for quadrilaterals to be parallelograms, i.e., one pair of opposite sides that are parallel and equal in length. If a quadruplet $[(\vec x_i,\vec x_ j),(\vec x_k, \vec x_l)]$ has passed through all the filters, the quadrilateral $P_i P_j P_k P_l$ is close to a parallelogram, and then another quadruplet $[(\vec x_i,\vec x_ k),(\vec x_j, \vec x_l)]$ also tends to pass through the filters. This effect enhances the number of quasars with even values of $s_i$. Any other zigzag features of this and the other histograms (Figures \ref{figure5}-\ref{figure8}) are not meaningful, but are merely stochastic.
 
\subsection{Some complexities due to global inhomogeneity}
We have seen that our method is so sensitive to real signal that even the limiting case can be detected. A slab space topology has been assumed there, but the method is valid in any spaces as will be shown in Appendix. However, in globally inhomogeneous spaces (see Table 1), situations become somewhat complicated. First, globally inhomogeneous spaces have specific, preferred coordinate systems by which their holonomies are defined, and our filters are also based on them. It means that there is no signal unless we use the correct coordinate system. Second, the shape of the Dirichlet domain depends on the observer's location in the universe, so the real signal may not be seen for some observers. We study these effects below.

To do this, we take, for example, a Hantzsche-Wendt space topology with $L_1=L_3=\frac{1}{\sqrt{2}}L=11.3$ Gpc, and $L_2 \rightarrow \infty$, since all the ``abnormal" effects due to global inhomogeneity can be seen in this space. With these parameters, four half-turn corkscrew  {motions} $T_z(\pm L_3)T_x(\pm L_1)O_z(\pi)$  {are} should be  detectable. 

As for spatial resolution, we have used $\varepsilon=5.8$ Mpc and $\vec \varepsilon=(7.3, 7.3, 7.3)$ in units of Mpc. These values are determined in a similar way to that in the previous section, but we have set the lower confidence level of 90\% and have taken the average values over 1000 observers randomly distributed in the fundamental cell. With the confidence level 99\%, spatial resolution becomes so high that the false signal dominates in some cases. Taking this into account, we proceed with the value of $90\%$.

\subsubsection{Choices of coordinate axes} 

In reality, we do not know the global directions of the  {Universe}, so we have to find them, if they exist, by changing our choices of the coordinate axes. This situation is analogous to the circles-in-the-sky method where we have to search for the centers of circles.

As long as the $z$-axis is correctly chosen, rotation about the $z$-axis does not affect anything, when we consider $T_z(\pm L/\sqrt{2})T_x(\pm L/\sqrt{2})O_z(\pi)$, so we only have to find the $z$-axis with some accuracy. We can roughly estimate the angular resolution $\Delta \theta$ by 
\begin{equation}
\frac{1}{2}L \Delta \theta \sim \varepsilon.
\end{equation}
Substituting $L =16$ Gpc and $\varepsilon=7.3$ Mpc, we derive $\Delta \theta \simeq 0.05 ^{\circ} \simeq 3'$. 

\begin{figure}[!htb]
\centering
\resizebox{\hsize}{!}{\includegraphics{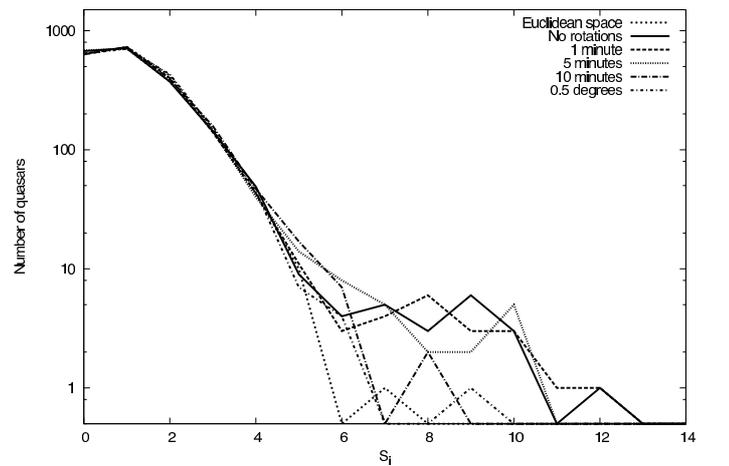}}
\caption{The $s_i$-histograms in the Hantzsche-Wendt space rotated about the $y$-axis by several angles. For comparison, the $s_i$-histogram is shown in simply connected space.}
\label{figure5}
\end{figure}

We have prepared two catalogs containing 1983 quasars in the Hantzsche-Wendt space in which the observer is at the center and in the  {simply connected  space}. The former catalog includes 13 pairs of ghosts.  { Before searching for quadruplets and applying our filters, we rotated the catalog} about the $y$-axis from the initial state, by $\Delta \theta=1', 5', 10', $ and $0.5^{\circ}$. The results are shown in Figure \ref{figure5}. It can be seen that the limit is in  between $5'$ and $10'$, roughly corresponding to the pre-estimated value $\Delta \theta \simeq 3'$. The area of the half sky is $A_{\mathrm{full sky}}/2=360^2/2\pi$ deg$^2$, so the number of trials needed to find the $z$-axis is
\begin{equation}
\frac{A_{\mathrm{full sky}}/2}{(\Delta \theta) ^2}=\frac{360^2/2\pi \times 60^2}{5^2} \simeq 3 \times 10^6,
\end{equation}
where we use $\Delta \theta=5'$. Simulation runs are mutually independent of each other, which takes about a few minutes with an ordinary personal computer. Therefore, a whole simulation can be accomplished faster by using many computers. No special computers are needed to clear up the grand mystery of the shape of our  {Universe}.

\subsubsection{Observer's location}
Next we consider cases in which the observer deviates from the center of the universe. We do not enter into the general theory of this effect, and simply show that the method works well even in these situations.  We have located the observer at several positions in the universe from which at least one pair of holonomies is detectable. 

For an observer at $\vec X=(X, Y, Z)$, the condition $|\vec X - \gamma \vec X| /2< r_2$ should be satisfied in order to  detect a holonomy $\gamma$ in a shell region of $r_1<r<r_2$. The lefthand side is the distance from the observer to the face of his Dirichlet domain, which corresponds to $\gamma$.  { Continuing with the same Hantzsche-Wendt space as an example, the lefthand side can be written explicitly as}
\begin{equation}
 \sqrt{\Bigl( X - \frac{L}{2\sqrt{2}} \Bigr)^2 + Y^2 + \frac{L^2}{8}} < r_2
 \end{equation}
 for $\gamma= T_z(\pm L/\sqrt{2})T_x(L/\sqrt{2})O_z(\pi)$, and 
 \begin{equation}
 \sqrt{\Bigl( X + \frac{L}{2\sqrt{2}} \Bigr)^2 + Y^2 + \frac{L^2}{8}} < r_2
\end{equation}
for $\gamma=T_z(\pm L/\sqrt{2})T_x(-L/\sqrt{2})O_z(\pi) $, where $L=16$ Gpc and $r_2=8.2$ Gpc. An observer satisfying each condition can detect each pair of  holonomies. It can be seen that the $Z$-location of the observer affects nothing. However, the bigger the $Y$-location becomes, the farther away all faces are located, hence the harder to detect all holonomies. The $X$-location has an effect of pushing two faces away, while drawing the other two, which is due to the translational direction that does not accord with the rotational axis. The last effect cannot be seen in spaces without half-turn corkscrew motion (type II) or glide reflection (type II).

\begin{figure}[!htb]
\centering
\resizebox{\hsize}{!}{\includegraphics{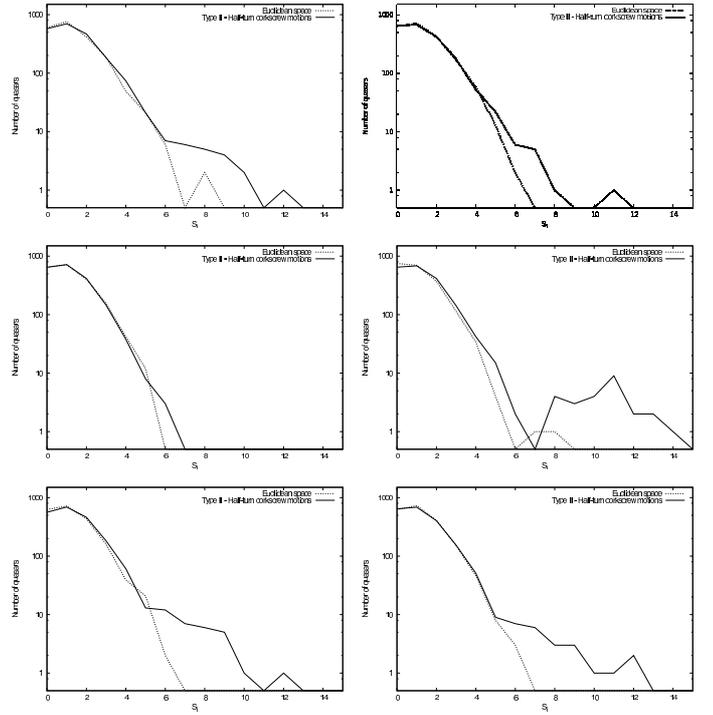}}
\caption{The dependence on the observer's location in the universe. Top left:(0, 0, 0), top right:(0, 0.5, 0), mid left:(0, 1, 0), mid right:(1, 0, 0), down left:(3, 0, 0), down right:(5, 0, 0).} 
\label{figure6} 
\end{figure}

According to these conditions, we chose six locations of $(X,Y,Z)=$ (0, 0, 0), (1, 0, 0), (3, 0, 0), (5, 0, 0), (0, 0.5, 0), and (0, 1, 0), in units of Gpc. We prepared catalogs of toy quasars seen from these observers, and applied our method to them. The coordinate axes were chosen correctly here. Results for these observers are given in Table \ref{table4} and Figure \ref{figure6}.

In this table, it can be seen that the signal gets weaker as the observer moves along the $y$-axis, since the faces get farther and the ghosts decrease in number. This is also seen in the histograms, where the hills constituted by ghosts disappear. It is necessary to use a larger shell region, if possible, to detect the ghosts.

As the observer moves along the $x$-axis, on the other hand, the signal and the hills remain like the initial one in which the observer is located at the center, since one pair of faces gets farther, but the others get closer. As long as the coordinate axes are chosen correctly, our method can detect holonomies that are close enough to the observer.  Our method is suited to obtaining a lower limit to the size of the  {Universe}.

\begin{table*}[!htb]
\centering
\caption{The results for different locations of the observer at $(X,Y,Z)$.} 
\begin{tabular}{c|c|c|c|c} \hline \hline
Location of & Number of & Number of & Topological & Topological \\
observer & quasars & ghosts & index $S_{\mathrm{mult}}$*  & index $S_{\mathrm{simp}}$* \\ \toprule
(0, 0, 0) & 2044 & 24 & $1.60 \times 10^{-10}$ & $1.44 \times 10^{-10}$ \\ 
(0, 0.5, 0) & 2014 & 18 & $1.50 \times 10^{-10}$ & $1.46 \times 10^{-10}$ \\
(0, 1, 0) & 1970 & 10 & $1.46 \times 10^{-10}$ & $1.47 \times 10^{-10}$ \\
(1, 0, 0) & 1972 & 24 & $1.63 \times 10^{-10}$ & $1.30 \times 10^{-10}$ \\
(3, 0, 0) & 2002 & 26 & $1.69 \times 10^{-10}$ & $1.46 \times 10^{-10}$ \\
(5, 0, 0) & 1970 & 26 & $1.59 \times 10^{-10}$ & $1.49 \times 10^{-10}$ \\ \bottomrule 
\end{tabular}
\\ \begin{footnotesize}*$S_{\mathrm{mult}}$ and $S_{\mathrm{simp}}$ represent the topological index defined in equation (3) for the Hantzsche-Wendt space and for the simply connected Euclidean space, respectively.\end{footnotesize}
\label{table4}
\end{table*}

\section{Summary and conclusions} 
\enlargethispage{0.5cm}
In this paper we have developed a cosmic crystallography method that significantly extends previous methods. A thin, shell-like part ($r_1<r<r_2$) of the full catalog is used, similar to the circles-in-the-sky method. This region should be as large as possible, but the comoving density of objects should be simultaneously high enough there. The quadruplets of these objects are filtered three times; separation, vectorial condition, and lifetime of objects. These filters drop the false stochastic ones, while keeping the real topological ones.
Flat spaces described by Euclidean geometry are assumed, and the second filter, vectorial condition, is specialized to detect their holonomies. This assumption is not fundamental, since similar filters for the holonomies in spherical or hyperbolic spaces can also be constructed.

The number of quadruplets having passed these filters is translated into an index of multiconnectedness of the universe, and then the objects are classified by $s_i$, the number of such quadruplets that include the object $P_i$ as one of their members. The $s_i$-histograms emphasize the topological ghosts: if ghost images really exist, they show some hills in the large-$s_i$ region.

We tested the method in catalogs of toy quasars with no observational uncertainties. We considered the severest cases for each type of holonomies, and any holonomies except for the targeted ones are assumed to be beyond the observed region. The translational distance of each holonomy is fixed with $L=16$ Gpc for which $8=(16/2)$ Gpc corresponds to $z \sim 5$, and the shell region is chosen as 7.8 Gpc$<r<$8.2 Gpc ($4.7 \lesssim z \lesssim 5.5$). In this situation, where the space is comparable to the observed region in size, existing crystallography method is no longer valid due to the contamination of false signal. It is found that our filters are able to eliminate such contamination, and the existence of topological ghosts is clearly recognized by the $s_i$-histograms. 
 { Our method considers isometries of flat spaces generically according to their type (translation, corkscrew motion,  or glide reflection), without requiring the specific manifold to be chosen \emph{a priori}. This is loosely analogous to the generic nature of the circles-in-the-sky method. Except for the case of translation, we have to search over several million possible orientations of the fundamental axes of the Universe, similar to the matched circles searching in the circles-in-the-sky method.} 
 
 For practical application of our method, more realistic simulations are necessary, and will be carried out in the next paper. They will include more realistic characteristics of quasars such as spatial correlations, activity cycles, and anisotropic morphology. Technical problems such as magnitude limits and selection biases will also be considered there.  {One possible selection effect relevant to our method, at least for translation, is clustering, which mimics a  topological signal as was discussed in Marecki et al. (2005). They constructed the $L_{\mathrm{selec}}$ filter, in the language of this paper, which removes a quadruplet $[(\vec x_i, \vec x_j),(\vec x_k, \vec x_l)]$ that is too compacted, i.e. satisfying $|\vec x_i - \vec x_j|, |\vec x_k- \vec x_l|<L_{\mathrm{selec}}$. This filter will also be useful for our method.}

Presently available data in the latest versions of the V\'eron-Cetty \& V\'eron quasar catalog (V\'elon-Cetty \& V\'eron 2010) and the Sloan Digital Sky Survey (SDSS) quasar catalog (Schneider et al. 2010) will be used to make more precise constraints, when compared with the previous constraints that ignore the lifetime of quasars.
Moreover, future observations that will detect hundreds of quasars with $z>6$ (e.g., the Joint Astrophysics Nascent Universe Satellite (JANUS)), will enable us to remove the disagreement in the observational constraints using CMB data; specifically, we will detect or exclude the cubic 3-torus topology with $L\simeq 3.8L_H$ mentioned by Aurich (2008).

\begin{acknowledgements}
{We thank Jeffrey Weeks and Adam Weeks Marano for beautiful figures of 17 multiconnected flat spaces. } We also thank T. Minezaki, T. Tsujimoto, T. Yamagata, Y. Sakata, T. Kakehata, and K. Hattori for useful discussions and suggestions. \end{acknowledgements}

\appendix
\section{Simulations for all types of holonomies}
\noindent We present the results for all types of holonomies here. 
The spatial resolutions used in these simulations are given in Table \ref{table5}. $\varepsilon '$ is defined as the maximum value of $\varepsilon _x$, $\varepsilon _y$, and $\varepsilon _z$, and we conservatively substitute $(\varepsilon ', \varepsilon ', \varepsilon ')$ for $\vec \varepsilon$.
Those for glide reflection (type II) were determined in the same way as for half-turn corkscrew motion (type II), while the others were determined in the same way as for parallel translation. The reason we did so is that, for the latter types of holonomies, the deviation of the observer's location from the center always pushes all the faces away, and makes it harder to detect them. We cannot observe any ghosts except for those that stand very close to the faces of our Dirichlet domain. Hence a relative cosmic time between a quasar $P_i$ and its ghost $\gamma (P_i)$ is very small, which leads to small positional uncertainties (integrated effect) and small spatial resolutions.

\begin{table}[!htb]
\centering
\caption{The spatial resolutions for each type of holonomies used here.} 
\begin{tabular}{c|c|c} \hline \hline
Holonomy & $\varepsilon$ (Mpc) & $\varepsilon'$ (Mpc)\\ \toprule
Translation & 3.7 & 6.0 \\
Half-turn corkscrew motion : I & 3.7 & 6.0 \\
Half-turn corkscrew motion: II & 5.8 & 7.3 \\
Quarter-turn corkscrew motion & 3.7 & 6.0 \\
Third-turn corkscrew motion &3.7& 6.0 \\
Sixth-turn corkscrew motion &3.7& 6.0 \\
Glide reflection : I & 3.7 & 6.0 \\
Glide reflection : II &5.3& 6.5 \\ \bottomrule
\end{tabular}
\label{table5}
\end{table}

\begin{figure}[!htb]
\centering
\resizebox{\hsize}{!}{\includegraphics{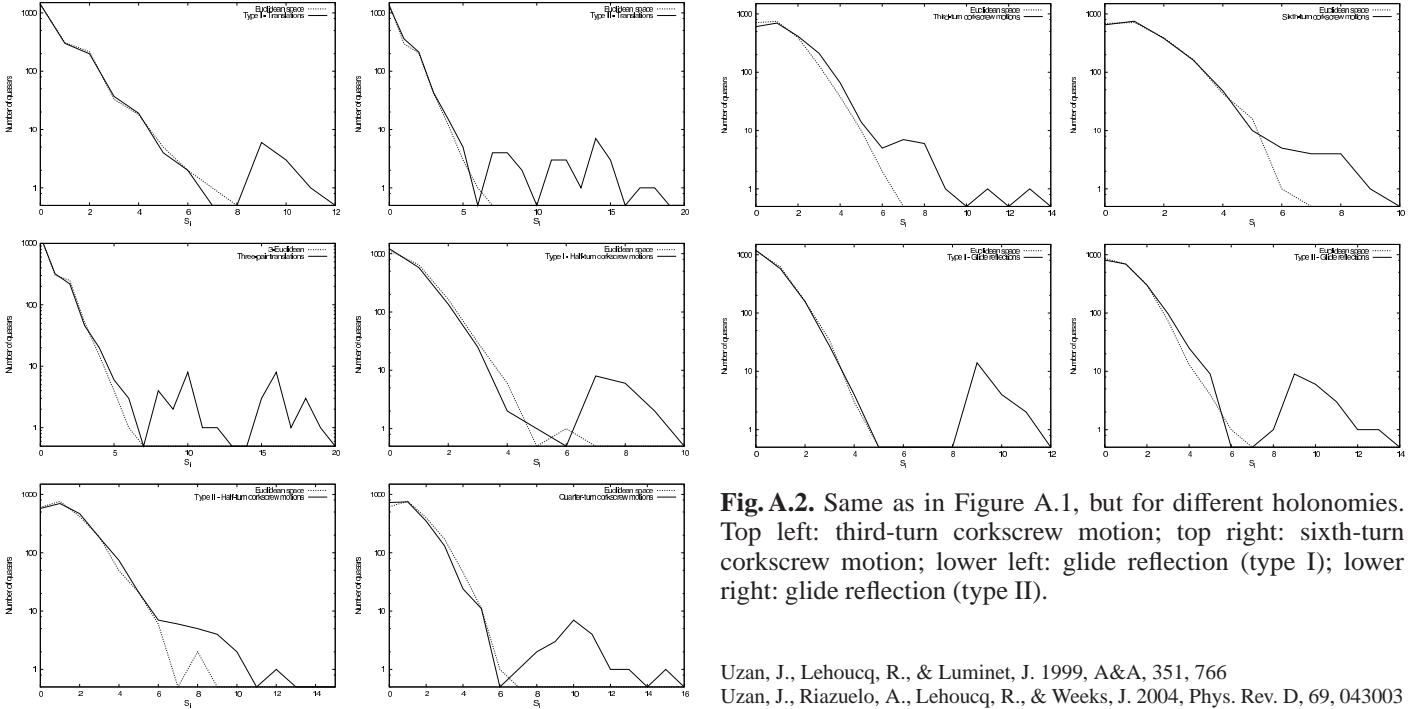}}
\caption{The $s_i$-histograms for each type of holonomy. Top left: one-pair translation; top right: two-pair translation; middle left: three-pair translation; middle right: half-turn corkscrew motion (type I); lower left: half-turn corkscrew motion (type II); lower right: quarter-turn corkscrew motion.} 
\label{figure7} 
\end{figure}

\begin{figure}[!htb]
\centering
\resizebox{\hsize}{!}{\includegraphics{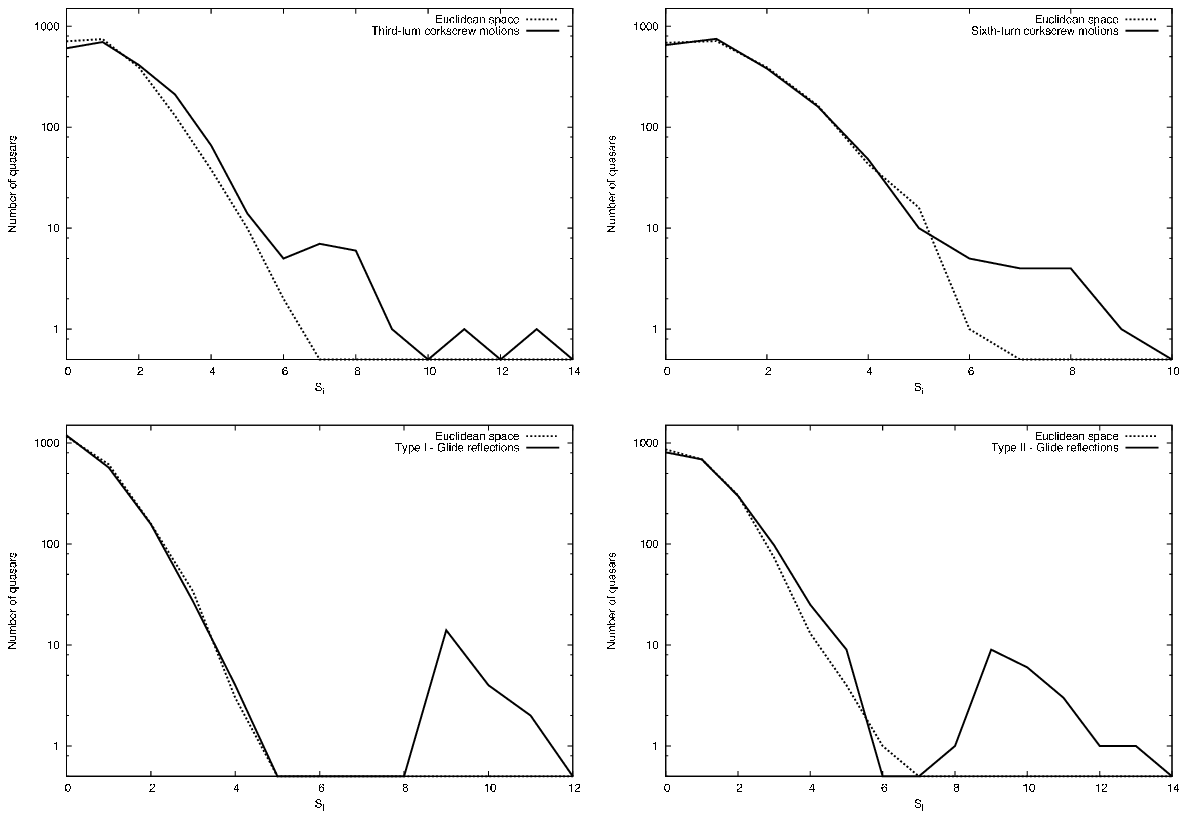}}
\caption{Same as in Figure \ref{figure7}, but for different holonomies. Top left: third-turn corkscrew motion; top right: sixth-turn corkscrew motion; lower left: glide reflection (type I); lower right: glide reflection (type II).} 
\label{figure8} 
\end{figure}

We located the observer at the center and used the correct coordinate axes. The situations in which these quantities deviate from them are discussed in section 5.2. The results are given in Table \ref{table6} and Figures \ref{figure7} and \ref{figure8}. Relatively strong signals for half-turn corkscrew motion (type II) and glide reflection (type II) stem from the large spatial resolution. The signals for $n$-th corkscrew motion for $n=4,3,$ and 6 are also strong, since we have to use two filters to detect $\gamma$ and $\gamma ^{-1}$. 

We can see that the hills always appear, which are constituted by topological ghosts. For a case where the clear distinction between the multiconnected space and the simply connected one is not seen, $s_i$-histograms are indispensable to distinguish them.
Our method is valid for all flat spaces. As mentioned in section 5.1, zigzag features seen in the hills are not notable. They are merely stochastic and different from calculation to calculation.

\begin{table*}[!htb]
\centering
\caption{The results for each type of holonomies.} 
\begin{small}
\begin{tabular}{c|c|c|c|c} \hline \hline
Holonomy type & Number of & Number of & Topological & Topological \\ 
& quasars & ghosts & index $S_{\mathrm{mult}}$ & index $S_{\mathrm{simp}}$ \\ \toprule
Translation : I & 1980 & 12 & $6.60 \times 10^{-11}$ & $6.24 \times 10^{-11}$ \\ 
Translation : II & 1964 & 28 & $9.01 \times 10^{-11}$ & $6.07 \times 10^{-11}$ \\
Translation : III & 1970 & 40 & $9.59 \times 10^{-11}$ & $6.95 \times 10^{-10}$ \\
Half-turn corkscrew motion : I & 1976 & 16 & $6.97 \times 10^{-11}$ & $7.21 \times 10^{-11}$ \\
Half-turn corkscrew motion : II & 2044 & 24 & $1.60 \times 10^{-10}$ & $1.44 \times 10^{-10}$ \\
Quarter-turn corkscrew motion & 2007 & 20 & $1.36 \times 10^{-10}$ & $1.43 \times 10^{-10}$ \\
Third-turn corkscrew motion & 2028 & 12 & $1.57 \times 10^{-10}$ & $1.27 \times 10^{-10}$ \\ 
Sixth-turn corkscrew motion & 2014 & 14 & $1.42 \times 10^{-10}$ & $1.37 \times 10^{-10}$ \\ 
Glide reflection : I & 1973 & 20 & $7.76 \times 10^{-11}$ & $6.94 \times 10^{-11}$ \\ 
Glide reflection : II & 1950 & 38 & $1.34 \times 10^{-10}$ & $1.11 \times 10^{-10}$ \\ \bottomrule
\end{tabular}
\end{small}
\label{table6}
\end{table*}

\nocite{*}
\bibliography{16521ref}

\end{document}